\documentclass[sn-mathphys, iicol]{sn-jnl}

\usepackage{pifont}

\usepackage{ulem}
\jyear{2023}%

\begin{document}

\title[RepSNet]{RepSNet: A Nucleus Instance Segmentation model based on Boundary Regression and Structural Re-parameterization}




\author[1]{\fnm{Shengchun} \sur{Xiong}}\email{xiongshengchun@foxmail.com}

\author*[1]{\fnm{Xiangru} \sur{Li}}\email{xiangru.li@gmail.com}

\author[1]{\fnm{Yunpeng} \sur{Zhong}}\email{1471881859@qq.com}

\author[2]{\fnm{Wanfen} \sur{Peng}}\email{wpf1500@sohu.com}

\affil*[1]{\orgdiv{School of Computer Science}, \orgname{South China Normal University}, \orgaddress{\street{West of Zhongshan Avenue}, \city{Guangzhou}, \postcode{510631}, \state{Guangdong}, \country{China}}}

\affil[2]{\orgname{	Signet Therapeutics}, \orgaddress{\street{289 Digital Peninsula, No. 2 Hongliu Road, Fubao Community, Fubao Street, Futian District}, \city{Shenzhen}, \postcode{518017}, \state{Guangdong}, \country{China}}}


\abstract{
Pathological diagnosis is the gold standard for tumor diagnosis, and nucleus instance segmentation is a key step in digital pathology analysis and pathological diagnosis. However, the computational efficiency of the model and the treatment of overlapping targets are the major challenges in the studies of this problem. To this end, a neural network model RepSNet was designed based on a nucleus boundary regression and a structural re-parameterization scheme for segmenting and classifying the nuclei in H\&E-stained histopathological images. First, RepSNet estimates the boundary position information (BPI) of the parent nucleus for each pixel. The BPI estimation incorporates the local information of the pixel and the contextual information of the parent nucleus. Then, the nucleus boundary is estimated by aggregating the BPIs from a series of pixels using a proposed boundary voting mechanism (BVM), and the instance segmentation results are computed from the estimated nucleus boundary using a connected component analysis procedure. The BVM intrinsically achieves a kind of synergistic belief enhancement among the BPIs from various pixels. Therefore, different from the methods available in literature that obtain nucleus boundaries based on a direct pixel recognition scheme, RepSNet computes its boundary decisions based on some guidances from macroscopic information using an integration mechanism. In addition, RepSNet employs a re-parametrizable encoder-decoder structure. This model can not only aggregate features from some receptive fields with various scales which helps segmentation accuracy improvement, but also reduce the parameter amount and computational burdens in the model inference phase through the structural re-parameterization technique. In the experimental comparisons and evaluations on the Lizard dataset, RepSNet demonstrated superior segmentation accuracy and inference speed compared to several typical benchmark models. The experimental code, dataset splitting configuration and the pre-trained model  were released at \url{https://github.com/luckyrz0/RepSNet
}.
}




\maketitle

\section{Introduction}
\label{sec:Introduction}

\begin{figure}[htbp]
    \centering
    \includegraphics[width=1\linewidth]{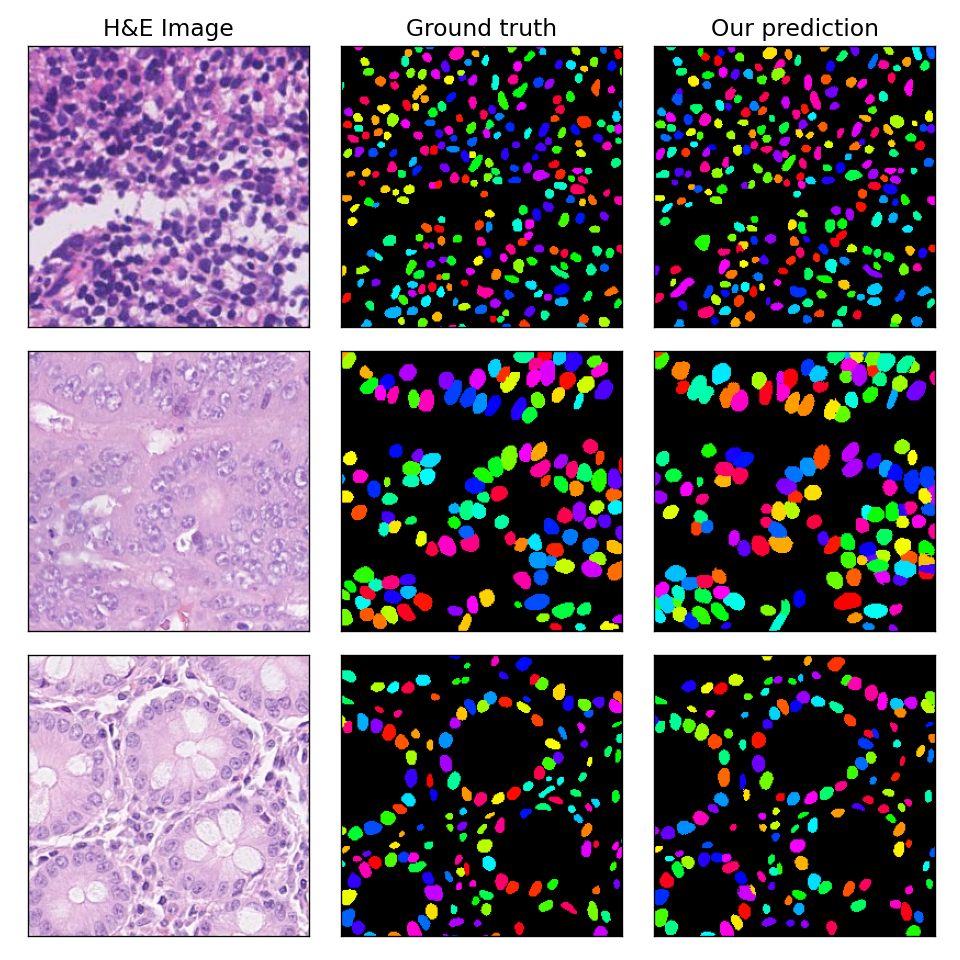}
    \caption{Challenges in the nucleus instance segmentation for histopathological images: a widespread adhesion of nuclei (in the cases in the first row), multiple species, shape variety, and low contrast between the nucleus foreground and the cytoplasmic background (in the cases in the second row). Three H\&E stained images, their instance segmentation ground truth, and predictions from the proposed method are repectively presented from left to right. In the images of the first column, nuclei are presented in dark color; In the images of the second column and the third column, various non-black colors are used to indicate different nuclei.
    }
    \label{fig:challenge}
\end{figure}

\begin{figure}[htbp]
    \centering
    \includegraphics[width=1\linewidth]{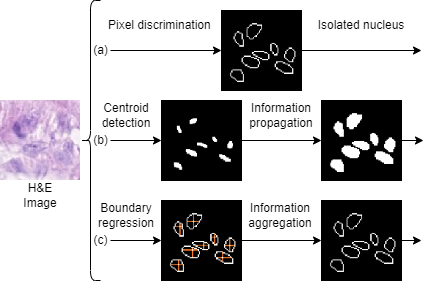}
    \caption{
    Three types of nucleus instance segmentation methods: (a) Pixel Discrimination (PD) method, (b) Nucleus Centroid Detection and Information Propagation (NCDIP) method, and (c) Nucleus Boundary Regression and Information Aggregation (NBRIA) method (\textbf{our proposed method}). The PD method obtains the nucleus foreground and boundary predictions based on some pixel-by-pixel classification schemes, and separates the overlapping nucleus instances based on the above-mentioned predictions. The NCDIP method first estimates the probability of each pixel belonging to one nucleus and extracts the centroid area (CA) of every nucleus using the predicted probability; Then, the NCDIP computes the outmost area for each nucleus and the corresponding instance segmentation result by propagating the information from the CA based on the local feature similarities of some neighbor pixels. NBRIA describes the nucleus instance segmentation as a boundary regression problem: NBRIA estimates the boundary position information (BPI) of the parent nucleus in four direction from each nucleus pixel using a proposed RepSNet, computes the nucleus boundary estimation by aggregating the BPIs from a series of pixels using a proposed boundary voting mechanism (BVM), and obtains the instance segmentation results by a connected component analysis procedure.
    }
    \label{fig:methods_difference}
\end{figure}

Pathological diagnosis is the gold standard of tumor diagnosis, and accurate nucleus instance segmentation is a key procedure in automatic pathological diagnosis. The characteristics of nucleus distribution and morphology (e.g. density, nucleocytoplasmic ratio, average size, and pleomorphism) are useful not only in assessing cancer grade but also in predicting therapeutic efficacy
\cite{Filipczuk2013Computer,sethi2015abstract}. The extraction of this kind features depends on the nucleus instance segmentation. However, pathological images are generally characterized by a widespread adhesion of nuclei, multiple species, shape variety, and low contrast between cytoplasmic background and nucleus foreground (Fig. \ref{fig:challenge})\cite{pan2022review}. These characteristics result in great difficulties to the nucleus instance segmentation.

Typical methods for nucleus instance segmentation in pathological images are the threshold segmentation and the watershed algorithm \cite{vincent1991watersheds,Yang2006Nuclei}. Such methods have poor segmentation accuracy on the cases in the presence of cohesive nuclei, or with low contrast between the cytoplasmic background and the nucleus foreground. To this end, a series of neural network models have been proposed for the nucleus instance segmentation in recent years (Fig. \ref{fig:methods_difference} (a) and (b)). Based on their principles, these methods can be roughly divided into two categories: Pixel Discrimination (PD) method\cite{chen2017DCAN,zhou2019CIA-Net,ZHAO2020Triple}, and Nucleus Centroid Detection \& Information Propagation (NCDIP) method\cite{Naylor2019DIST,GRAHAM2019HoverNet,Yao2021PointNuNetSM}. However, the nucleus boundary identification performance of these two types of methods is limited by the exploitation of contextual information at the nucleus level. This limitation makes it difficult for these methods to accurately make decision in some areas with blurred nucleus boundaries and cohesive nuclei (Fig.\ref{fig:Qualitative_analysis2}, more analyses are made in section \ref{sec:RelatedWork}).

To this end, a novel scheme named Nucleus Boundary Regression and Information Aggregation (NBRIA) is proposed for nucleus instance segmentation and classification in H\&E-stained histopathological images.
Inspired by the NCDIP approach, NBRIA reformulates the nucleus instance segmentation task as an object boundary regression combined with a boundary voting procedure. First, the NBRIA scheme estimates boundary position information (BPI) in four directions from each pixel in the inner region of the parent nucleus. Then, a Boundary Voting Mechanism (BVM) is proposed to aggregate the BPIs which are used in obtaining more accurate nucleus boundaries. In estimating a boundary pixel based on the BVM, NBRIA uses the BPI information from a series of pixels of a nucleus. Therefore, this mechanism improves the reliability of the estimation and the robustness of the segmentation algorithm.

As mentioned above, nuclei are characterized by their shape variety, wide scale range, and low contrast between cytoplasmic background and nucleus foreground etc. These characteristics make it necessary to establish an instance segmentation model with complex feature representation capability, multi-scale feature extraction ability and weak feature sensitivity.
 To this end, this work designed a nucleus instance segmentation network called RepSNet based on the NBRIA scheme. This network has an encoder-decoder structure. In the encoder, RepSNet uses a re-parameterizable RepVgg Unit \cite{Ding2021RepVGG}. This configuration helps systematically extract the features from various receptive fields.
 In the training phase, RepVgg Unit extracts features with various scales through three independent branches. In the inference phase, RepSNet reduces the parameter amount and computation burdens of the model by reparameterizing the three branches into one branch. In the decoder, we also introduce the structural re-parameterization technique into the deconvolution layer to further enhance the multi-scale perception capability of the model.

Moreover, pathological images are characterized by a widespread adhesion, the existence of a large number of nuclei, a low contrast between nucleus foreground and cytoplasmic background. These characteristics result in some possible boundary annotation errors in the training data. To reduce the negative impacts from these misannotations on model learning, this work proposes a loss function $L_{nb}$ based on some nucleus boundary isoheights. This loss function can adaptively penalize the deviations of boundary estimation. And the penalization adaptivity makes RepSNet well suppress the interferences from anomalous nucleus boundary labeling.

To evaluate the effectivity of NBRIA(RepSNet), this work conducted a series of comparing experiments on the Lizard dataset \cite{Graham2021Lizard}. This dataset is used in the 2022 Colon Nuclei Identification and Counting (CoNIC) Challenge \cite {graham2021conic,CoNiC2023arXiv}. The Lizard dataset consists of a series of pathological images of colonic tissues, and are constructed based on GlaS \cite{Korsuk2017glas}, CRAG \cite{GRAHAM2019MILD-Net}, CoNSeP \cite{GRAHAM2019HoverNet}, DigestPath \cite{DigestPath2022}, PanNuke \cite{PanNuke2019,gamper2020multi} and TCGA \cite{grossman2016toward}. The dataset is characterized by a large number of labeled instances and its diversity of targets. In this dataset, there are approximately 500,000 labeled nuclei of six categories: neutrophil, epithelial, lymphocyte, plasma, eosinophil, and connective. Extensive experiments on the Lizard dataset show that the proposed NBRIA(RepSNet) has excellent instance segmentation performance and high computational efficiency compared to some benchmark models. On the test set, the mPQ metric of NBRIA(RepSNet) reached 0.5633. Compared with the CoNIC SOTA StarDist model, NBRIA(RepSNet) improves it by 0.0161. This result indicates that NBRIA(RepSNet) has a significant performance improvement on the Lizard dataset with a large number of nuclei.

The main contributions of this work are as follows:
\begin{itemize}
    \item A novel nucleus instance segmentation scheme based on nucleus boundary regression and information aggregation (NBRIA) is proposed. This scheme discriminates nucleus boundaries from other pixels by aggregating a series of boundary position estimations from some nucleus pixels. This is a process of multi-scale feature exploitation, fusion, and collaborative inference,  rather than a direct pixel-by-pixel classification procedure. This approach improves the discrimination performance on the boundaries of cohesive nuclei.
    \item A full re-parameterizable encoder-decoder network RepSNet is designed based on the NBRIA scheme. This network is implemented based on some structural re-parameterization techniques which improve the model's efficiency and feature extraction performance from some receptive fields with various scales without increasing the computational complexity during inference. The NBRIA(RepSNet) achieves an mPQ of 0.5633 on the Lizard dataset and can process 10 pathological images with $256\times256$ pixels per second.
    \item A loss function is proposed based on boundary isoheights. This loss function enhances the model's adaptability to the potential misannotations in training dataset by adaptively penalizing the boundary estimations with various deviations from boundary annotations.
\end{itemize}

\section{Related works}
\label{sec:RelatedWork}

The available deep learning methods for pathological image instance segmentation in literature can be can be divided into two categories: pixel discrimination (PD) method, and nucleus centroid detection and information propagation (NCDIP) method (Fig.\ref{fig:methods_difference}).
This section summarizes and analyzes these two types of approaches.

\subsection{Pixel Discrimination (PD) method}
\label{sec:PD_method}

The PD method describes the nucleus instance segmentation task as a pixel classification problem. To determine whether each pixel is a nucleus pixel, a nucleus boundary pixel or a background pixel, such methods generally utilize a Convolutional Neural Network with an encoder-decoder structure \cite{Olaf2015UNet, Zhou2018UNet++}. The models of this kind classify all pixels into two fundamental types using their local features: nucleus pixels, or background pixels. To segment overlapping nucleus instances, the PD methods are further augmented with an estimator for nucleus boundary properties \cite{chen2017DCAN, Kumar2017dataset}; These schemes are usually equipped with three branches in their decoder end respectively for determining the nucleus pixels and the nucleus types (NPNT branch), boundary pixels (BoP branch), and background pixels (BaP branch). Recently, several scholars attempted to establish some communication channels between the NPNT branch and BoP branch \cite{zhou2019CIA-Net,ZHAO2020Triple}.

The typical representatives of the PD method include DCAN \cite{chen2017DCAN},CIA-Net \cite{zhou2019CIA-Net},Triple U-Net \cite{ZHAO2020Triple}. These methods all first obtain nucleus foregrounds and nucleus boundaries by a direct pixel classification approach, and then compute the overlapping nucleus instances using the extracted nucleus boundaries. In the cases of uneven H\&E staining or uneven slice thickness, it is possible that there exist weak nucleus boundaries or blurred boundaries in the pathological images. However, a PD method recognizes nucleus boundaries only based on some local information of some nucleus pixels, but not global information or the fusion of these two kind information. And it is clearly insufficient in terms of exploiting the contextual, global information of nuclei. This kind insufficiency results in some nucleus boundary misdetections in the PD method processing results (Fig. \ref{fig:Qualitative_analysis2}).

Different from the PD method, the proposed NBRIA(RepSNet) does not directly determine nucleus boundary properties pixel-by-pixel. Instead, it estimates the boundary position information (BPI) of the parent nucleus for each pixel, and gives an estimate for the nucleus boundary by fusing the BPI of a series of pixels using a voting scheme. NBRIA estimates the boundary for a nucleus by utilizing the information from every pixel, and exploits the complementary information between these pixels in this procedure. Therefore, the nucleus boundary estimations from NBRIA are more reliable than those from the PD method which determines the boundaries simply using a pixel-by-pixel classification scheme.

\subsection{Nucleus Centroid Detection and Information Propagation (NCDIP) method}
\label{sec:NCDIP_method}

To overcome the limitations of PD method in boundary detection, the NCDIP method adopts a strategy based on the extraction and propagation of key information. In each nucleus, a subset of its pixels can be detected more reliably and easily than others. This subset of pixels is referred to as the centroid area (CA) of a nucleus. Based on the detected CAs, NCDIP can estimate the fundamental informations, such as locations, quantity, and identities of nucleus instances. Nucleus identity refers to the distinction between various nuclei and their types. Based on some similarities between adjacent pixels computed from local features, NCDIP propagates the key information from a CA to the peripheral pixels of the parent nucleus, and obtains the instance segmentation results and nucleus boundary prediction.

Typical representatives of NCDIP method include DIST \cite{Naylor2019DIST}, Hover-Net \cite{GRAHAM2019HoverNet}, and PointNu-Net \cite{Yao2021PointNuNetSM}. The NCDIP method estimates the probability of each pixel belonging to a nucleus based on the shortest distance between the pixel and its parent nucleus’s boundary or centroid. NCDIP computes the CA detection results based on these probability estimates. Consequently, the NCDIP method can exploit the global, contextual information about a nucleus better than the PD method. Therefore, in some regions with overlapping nuclei, the NCDIP method exhibits superior segmentation capabilities compared to the PD method. However, the performance of the NCDIP method largely depends on the  completeness of CA detection. Especially, it is shown that the CAs of small nuclei might fail to be detected (Fig. \ref{fig:Qualitative_analysis2}). This characteristic result in some failure detection for small nuclei in the NCDIP method. In addition, the NCDIP method only uses some local information of adjacent pixels in propagating the key information of a nucleus from a CA to the peripheral pixels of the parent nucleus. Therefore, similar with the PD method, the NCDIP method also fails to accurately recognize the boundaries of the overlapping nuclei.

Although both NCDIP and the proposed NBRIA describe the instance segmentation problem as a nucleus key information extraction problem based on a regression scheme, NBRIA does not focus on using key information to detect key regions. Instead, it first estimates the nucleus boundaries by aggregating the predicted boundary location information from each nucleus pixel, and then obtains instance segmentation results based on connectivity analysis. Since this aggregation process incorporates some information from the entire internal region of a nucleus, NBRIA can maintain good performance in segmenting small nucleus. Compared with the NCDIP method, which relies on key region detection, NBRIA comprehensively utilizes the information in an image.  Therefore, NBRIA reduces the risk of missing small nuclei and improves the robustness of instance segmentation.

\section{NBRIA scheme and RepSNet network for nucleus instance segmentation}
\label{sec:Method}

In this paper, a novel nucleus instance segmentation scheme is proposed based on Nucleus Boundary Regression and Information Aggregation (NBRIA). This scheme estimates the boundary position information (BPI) from each pixel to discriminate the nucleus boundary pixels based on a voting procedure (Boundary Voting Mechanism, BVM). Compared with the available PD and NCDIP methods, the NBRIA scheme improves the ability of segmenting the overlapping nucleus instances. In the NBRIA, a RepSNet network with an encoder-decoder structure is designed to maximize the exploiting of the nucleus boundary information and nuclear type information. The network enhances the ability of extracting features from some receptive fields with various scales by virtue of a structural re-parameterization technique.

In the following subsections, we firstly introduce the fundamentals and implementation process of the NBRIA (section \ref{sec:NBRIA_scheme}). Secondly, we present the structure of RepSNet (section \ref{sec:Network}), including the re-parameterizable RepVGG Unit \cite{Ding2021RepVGG} and the re-parameterizable upsampling module RepUpsample. Thirdly, we describe the implementation of the boundary voting mechanism (BVM) (section \ref{subsubsec:bvm}). Finally, we design a loss function for learning RepSNet (section \ref{sec:Loss}), and describe the connectivity analysis procedure and nucleus instance classification in section \ref{subsec:connectivityAnalysisNucleusClassification}.

\subsection{NBRIA scheme for nucleus instance segmentation}
\label{sec:NBRIA_scheme}

\begin{figure*}[htbp]
    \centering
    \includegraphics[width=1\linewidth]{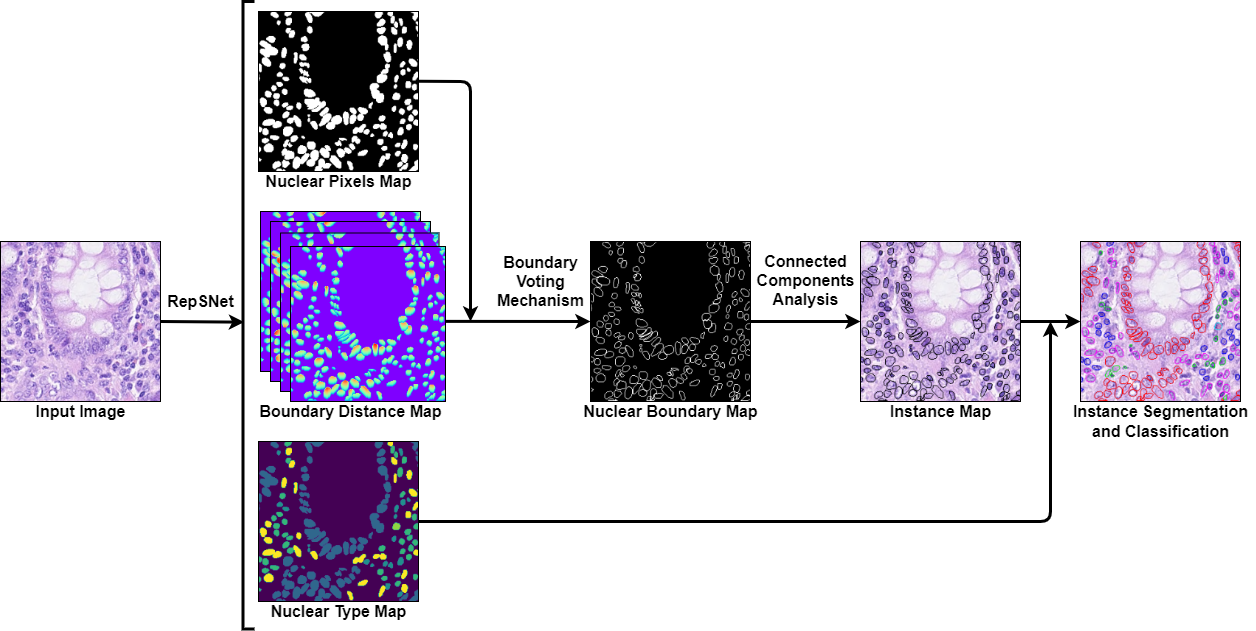}
    \caption{ A flowchart of the proposed NBRIA for nucleus instance segmentation and recognition. For an input image, the proposed RepSNet computes three direct outputs: a nucleus pixel (NP) feature map, a nucleus type (NT) feature map, and a boundary distance (BD) feature map. The BD feature map is a matrix consisting of the distances from each pixel to the boundary of its parent nucleus in four directions. Furtherly, RepSNet estimates a nucleus boundary (NB) feature map from the NP feature map and BD feature map using a proposed Boundary Voting Mechanism (BVM, section \ref{subsubsec:bvm}). This BVM procedure fuses the boundary position information (BPI) estimated from each pixel of a nucleus. Therefore, RepSNet systematically exploits local information and global contextual information in determining the nucleus boundary. Then, the NBRIA computes a nucleus instance (NI) feature map from the NB feature map using a connectivity analysis procedure. Finally, the type of each instance is determined by the most frequent occurring type in that instance area.
    }
    \label{fig:NBRIA_scheme}
\end{figure*}

The NBRIA deals with the nucleus instance segmentation problem as follows: 1) estimates the BPIs from each pixel, 2) determines the nucleus boundary by aggregating the BPI information using a voting mechanism, 3) refines the boundary estimation by performing connectivity analysis. The available PD method in literature implements instance segmentation by determining the type for each pixel based on the local features around it (Fig. \ref{fig:methods_difference}, Sec. \ref{sec:PD_method}). The available NCDIP method computes the nucleus instance segmentation result by detecting the CA for each nucleus and propagating some information from the detected CA to the other pixels of the parent nucleus based on some local feature similarity between two adjacent pixels (Fig. \ref{fig:methods_difference}, Sec. \ref{sec:NCDIP_method}). Although NCDIP exploits both local features and global features in detecting CAs, this method depends only on some local information of nucleus pixels in the process of belief propagation. The proposed NBRIA also extracts both some local features and the contextual global features in estimating the boundary position of the parent nucleus from each pixel. Especially, NBRIA exploits the synergistic belief-enhancement features between various pixels in aggregating their BPI estimations using a voting mechanism and connectivity analysis. Synergistic belief-enhancement refers to the following statistical mechanism: although there exist some deviations on many of the boundary position information (BPI) estimations, the BPI estimations with little deviations show more consistency than the others in boundary voting, and this consistency help us compute accurate nucleus boundary prediction from their votes; The votes from the BPI estimations with large deviations usually show much inconsistency \& dispersion and can be filtered out as noises. Therefore, the Boundary Voting Mechanism (BVM) and connectivity analysis improve the robustness and fault tolerance in the NBRIA.

A flowchart of the NBRIA scheme is presented in Fig. \ref{fig:NBRIA_scheme}. After inputting an H\&E-stained pathology image into NBRIA, it is processed by RepSNet network. The RepSNet has an encoder-decoder structure with two branches in its decoder end. Two decoder branches respectively compute (a nucleus pixel (NP) feature map, a nucleus type (NT) feature map), and a nucleus boundary distance (Boundary Distance, BD) feature map. The NP feature map and the NT feature map respectively consist of the estimated identity (a nucleus pixel or a background pixel) of each pixel and the type of the pixel's parent nucleus. The BD feature map consists of the distances from each image pixel to the boundary of its parent nucleus in four directions. Specifically, the BD map is a matrix with the same size as the input image, and each element of this matrix represents the boundary distance information between the corresponding pixel and the boundary of the parent nucleus. Suppose $p$ is a pixel, through $p$ we makes two straight lines parallel respectively to X axis and Y axis. Let A, B, C and D respectively represent four intersections of the two lines and the boundary of the pixel's parent nucleus. Then, the boundary distance information of the pixel $p$ is a four-dimensional vector, and the four components of this vector represent the Euclidean distance from $p$ to A, B, C, and D, respectively.

 To obtain accurate nucleus instance masks, we propose a post-processing pipeline based on the idea of BD information aggregation. The pipeline first converts the estimated BD feature map into a nucleus boundary (NB) feature map using a Boundary Voting Mechanism (section \ref{subsubsec:bvm}). The NB feature map is a mask map with the same size as the input image. In this mask map,  1 and 0 respectively indicate that the corresponding position is a nucleus boundary pixel or other pixels. Then, the NBRIA instantiates each nucleus and enhances the detected boundary  by a connectivity analysis based on a nearest neighbor classifer. The computed results in this procedure are referred to as a nucleus instance (NI) feature map (Fig. \ref{fig:NBRIA_scheme}). Finally, the NBRIA determines the nucleus type by computing the pixel voting results from the NT feature map.

\subsection{RepSNet}

\subsubsection{Network Architecture and Structural Re-parameterization Techniques}
\label{sec:Network}

\begin{figure*}[htbp]
    \centering
    \includegraphics[width=1\linewidth]{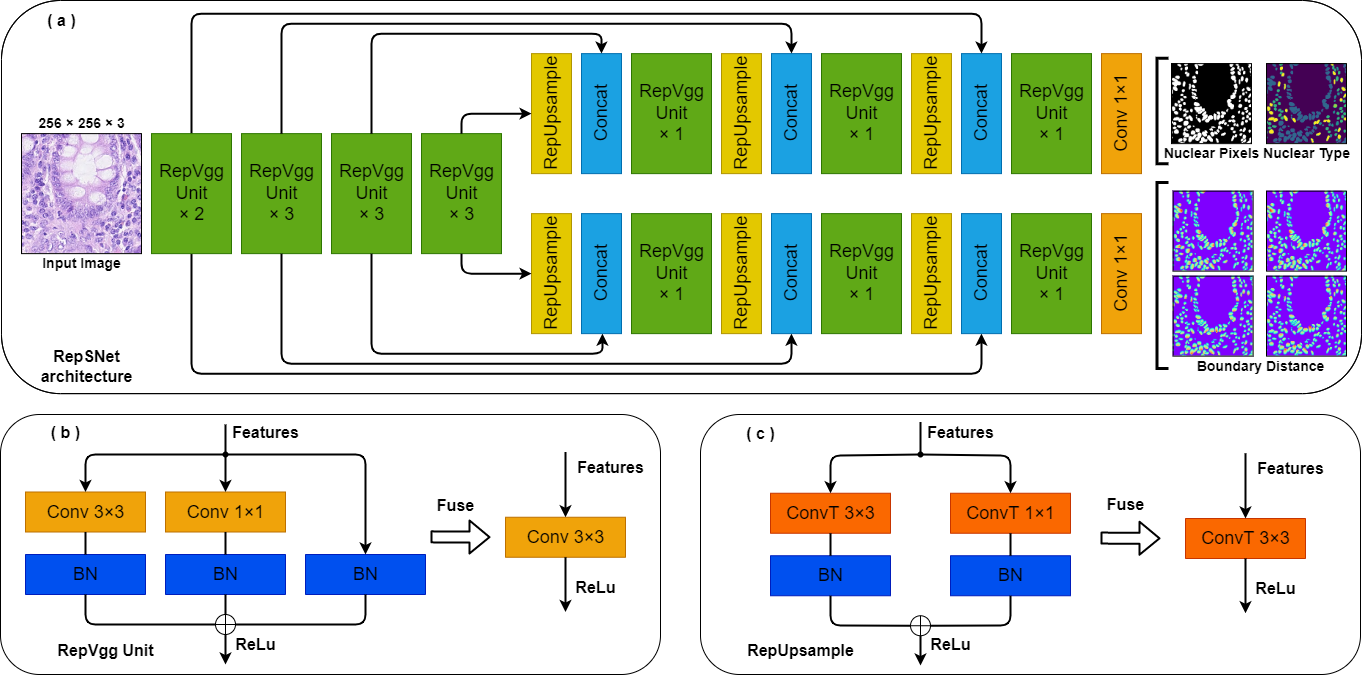}
    \caption{
    The network architecture of the proposed RepSNet.
    (a) RepSNet is a U-Net-like re-parameterizable neural network model with a shared encoder, and two structurally identical and mutually independent decoders.
    One decoder is used to determine whether a pixel comes from a nucleus, and which category of the parent nucleus (seven categories of nuclei in the experiments of this work). The other decoder is a regression branch for estimating the distances from each pixel to the boundary of the parent nucleus in four directions (up, down, left, and right).
    (b) RepVgg Unit \cite{Ding2021RepVGG} is a feature extraction module with structural re-parameterization functions. In the training phase, the RepVgg Unit has three convolutional branches with different-size receptive fields to improve the sensitiveness of the model to multi-scale features.
    In the inference phase, three branches of the RepVgg Unit are fused into one convolutional branch by a structural re-parameterization technique to reduce the parameter amount and the computation burdens of the model.
    (c) RepUpsample is a proposed upsampling module with a structural re-parameterization technique. In literature, the structural re-parameterization technique is used in some downsampling procedures.
    }
    \label{fig:RepSNet_Architecture}
\end{figure*}

Inspired by the encoder-decoder structure in U-Net, RepSNet uses a shared encoder and two independent decoders to implement the nucleus instance segmentation and classification (Fig. \ref{fig:RepSNet_Architecture}(a)). One of the two decoder branches is to predict the NP and NT feature maps, and another decoder branch is used to estimate the BD feature map. The dual-decoder structure enables RepSNet to improve the reliability of the instance segmentation by exploiting personalized features for two supervised learning tasks. The shared encoder in RepSNet not only reduces the number of parameters of the model to some extent, but also improves the instance segmentation accuracy of the model by exploiting the complementary information between two supervised learning tasks.

The model with an encoder-decoder structure improves the instance segmentation fineness and accuracy to a certain extent with the advantages of comprehensively exploiting and fusing the deep features and shallow features. However, the nucleus instance segmentation problem is characterized by large target shape differences\&diversity, high density, and even serious adhesio. Some experimental results also indicate that it is difficult to deal with the above-mentioned challenges simply by fusing some deep features and shallow features, especially in the cases with blurred boundaries or small nuclei (Fig. \ref{fig:Qualitative_analysis1}). To enhance the model's capability of exploiting features from some nuclei with various sizes and shapes, therefore, many researchers have tried to add some multi-scale perception mechanisms to the model.
 For example, DCAN \cite{chen2017DCAN} is equipped with the multi-scale perception capability by exploiting some multi-level contextual features, and improves learning performance by adding several weighted auxiliary classifiers.
 However, the multiple auxiliary classifiers in DCAN only independently optimize the features with various scales, but do not deeply fuse them. Vu et al.\cite{vu2019DRAN} proposed a scheme to improve segmentation performance by combining input images at multiple scales. This approach requires the model to process the input images with various sizes simultaneously. This characteristic result in high computational burdens. Recently, Yao et al. \cite{Yao2021PointNuNetSM} proposed a Joint Pyramid Fusion Module (JPFM). This module extracts some features from a series of receptive fields with various scales using some dilated convolutions and fuses these feature using  a cascading scheme. Although JPFM helps increase the exploitation of multi-scale features, it leads to a huge number of parameters and  high computational burdens.

To control computational complexity and increase the model's sensitivity to multi-scale features, we introduced a large number of RepVGG Units \cite{Ding2021RepVGG} into RepSNet (Fig. \ref{fig:RepSNet_Architecture}(b)). The RepVGG Unit consists of three parallel branches in the training phase, namely a $3\times3$ convolution branch, a $1\times1$ convolution branch, and a residual learning branch. These three parallel branches force the model to extract the nucleus features from some receptive fields with various scales during the training phase. Therefore, RepSNet can adapt to the instance segmentation of the nuclei with various sizes. As mentioned above, in order to improve the multi-scale target instance segmentation ability of the model, JPFM adopts a strategy of simply cascading multiple parallel branches \cite{Yao2021PointNuNetSM} in both the training and application stages. This strategy leads to an evident increase in the number of model parameters and computational burdens. To overcome this limitation of JPFM, the RepVgg Unit used in this work fuses the three parallel branches into a $3 \times 3$ convolution branch based on the parameter re-parameterizaion technique in its inference stage. The re-parameterized convolutional branches not only reduce the parameter amount of the model, but also maintain the multi-scale object-sensing capability learned in the multiple branch configuration during the training phase.

To further improve the sensitiveness of the network to multi-scale features, this work also introduces the structural re-parameterization technique into the upsampling module of RepSNet. The upsampling module with a structural re-parameterization technique is referred to as RepUpsample (Fig. \ref{fig:RepSNet_Architecture}(c)). Although there exist some similarities on their structure between RepUpsample and RepVgg Unit \cite{Ding2021RepVGG}, the former is a novel re-parameterizable unit designed for a deconvolution scenario. During training phase, RepUpsample adopts a dual-branch structure: a $3\times3$ deconvolution branch and a $1\times1$ deconvolution branch. During inference phase, these two branches are reconstructed into a $3\times3$ deconvolution unit. Different from the traditional deconvolution upsampling based on a single $2 \times2$ kernel, RepUpsample conducts a more detailed instance segmentation by fusing the information from two receptive fields with different scales. Note that to ensure that the input features are correctly upsampled by a factor of 2, the deconvolution unit should use a configuration with stride=2 and output\_padding=1.

To sum up, RepSNet is a thoroughgoing re-parameterization model with  a large number of RepVgg Units in its encoder and a series of RepUpsample Units in its two decoder branches. During training phase, each RepVgg/RepUpsample Unit enhances RepSNet's sensitivities to multi-scale nuclei using three/two convolutional/deconvolutional branches with various-size receptive fields.
 During inference phase, each RepVgg/RepUpsample Unit completely copies the learned results using only one convolutional/deconvolutional branch. Therefore, this thoroughgoing structural re-parameterization technique significantly improves the computational efficiency and the sensitivities of the proposed RepSNet to multi-scale nuclei.

It is necessary to note that in training phase, a RepSNet with a single {$3\times3$} convolutional branch theoretically should also be able to learn the features sensitive to multi-scale nuclei. However, some experimental results show that the model performance in this case exhibits a significant degradation (Table \ref{tab:Ablation}). The cause is that there exist some masking effects between the features from the nuclei with various scales in RepSNet with a single {$3\times3$} convolutional branch. The masking effect refers to the following phenomenon: in a dataset, if the number of the nuclei with one scale is far more than that of the nuclei with another scale, the latter can be overlooked in a high probability by a learning system. On the contrary, in a learning system equipped with several parallel convolutional branches with various scales, each convolutional branch only need to focus on the features with a specific scale. In this case, the learning task is relatively easy.  Thus, in the training phase RepSNet makes it capable of extracting nucleus features with various scales by virtue of a large number of parallel branches; In the inference phase, RepSNet distills and fuses the  learned knowledges from these parallel branches using a re-parameterization techniques. The re-parameterization reduces the parameter amount and computational burden of the model.

\subsubsection{Belief Aggregation and Synergistic Enhancement: Boundary Voting Mechanism}
\label{subsubsec:bvm}

To compute complete and accurate nucleus instance segmentation results, an additional post-processing pipeline is usually required to further process the output of a network. For example, Kumar et al \cite{Kumar2017dataset} proposed a post-processing pipeline based on region growing to process the nucleus boundary feature maps computed from a network. {DCAN}\cite{chen2017DCAN} investigated a post-processing pipeline for complete nucleus instance segmentation results based on the watershed algorithm to fuse the object probability map and object contour probability map computed from a segmentation network.
Hover-Net\cite{GRAHAM2019HoverNet} first estimated the horizontal and vertical distance maps from each pixel to the centroid of the parent nucleus using a neural network, and then similarly post-processed it using the watershed algorithm. StarDist\cite{StarDist2018} first estimate a candidate nucleus boundary from each nucleus pixel, and then select one of the candidate boundaries as the final boundary estimation by a Non-Maximum Suppression (NMS) scheme. However, the above-mentioned post-processing pipelines all computed their nucleus instance segmentation based on the similarity between pixels using their local features, but do not integrate these local features with the global information of a nucleus.

In contrast to the aforementioned methods, we propose a Boundary Voting Mechanism (BVM) as a post-processing pipeline for RepSNet predictions. This mechanism aggregates the RepSNet predictions for nucleus boundary in each row/column of BD feature map by voting. In applications, the position of nucleus boundary can be determined synergistically by a series of pixels in its inner region. And experimental results also show that we usually can obtain excellent boundary estimations from a large fraction of nucleus pixels.  Therefore, the BVM algorithm can compute accurate final boundary predictions, improve the robustness of the model while suppressing some erroneous boundary position predictions. The proposed BVM mechanism is described in the algorithm \ref{alg:BVM}.

\begin{algorithm}
\caption{Boundary Voting Mechanism (BVM)}\label{alg:BVM}
\begin{algorithmic}[1]

\Require Nucleus boundary distance feature map {$\mathrm{BD} \in \mathbb{R} ^{H \times W \times 4}$}, nucleus foreground pixel feature map {$\mathrm{NP} \in \mathbb{R} ^{H \times W}$}, boundary threshold {$e_t$}.
\Ensure  Nucleus boundary feature map {$\mathrm{NB} \in \mathbb{R}^{H \times W} $}

\State {Initialization: $\mathrm{NB} \Leftarrow \mathbf{0}$}
\State Define a variable {$m \in \mathbb{R} ^{N_t \times 2}$} representing the indexes for the elements with {$\rm NP>0$}, where {$N_t$} is the number of nucleus foreground pixels in {$\rm NP$}.
\State Use index $m$ to obtain the relative distances {$bd_r \in \mathbb{R}^{N_t \times 4}$} from the foreground pixels of the nucleus to the boundary in all four directions at each position.
\State Convert the relative distance {$bd_r$} to the absolute pixel position of the boundary {$bd_a \in \mathbb{R} ^{4N_t \times 2}$}.
\For{$i=1$ to $4N_t$}
    \State {$\mathrm{NB}[bd_a[i]] = \mathrm{NB}[bd_a[i]] + 1$}
\EndFor \\
\Return {$\mathrm{NB} > e_t$}
\end{algorithmic}
\end{algorithm}

First, we obtain the indexes $\{m_i\}$ of the nucleus foreground pixels from the NP feature map. In the  BD feature map, we extract the nucleus boundary relative position predictions,  $\{ bd_r \}$, estimated from the pixels with indexes $\{m_i\}$. Then, we transform the relative distances $\{ bd _ r \}$ to the absolute pixel positions $\{ bd_a\}$ of the nucleus boundaries. Since each pixel within a nucleus predicts the nucleus boundary, it is inevitable that the same position of the nucleus boundary will be repeatedly predicted by multiple pixels. Therefore, we can count the number of repeated estimations for each nucleus boundary pixel and take the estimation with repetitions greater than a threshold $e_t$ ( set to 3 in this paper) as the final estimation for nucleus boundary.

\subsubsection{Loss function}
\label{sec:Loss}

\begin{figure}[tbp]
    \centering
    \includegraphics[width=1\linewidth]{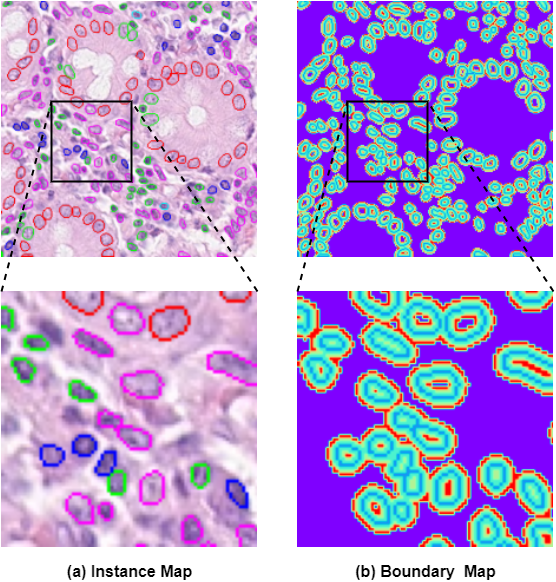}
    \caption{ Two sketch maps for computing the nucleus boundary loss ($L_{nb}$).
    (a) A sketch map of nucleus instance segmentation. (b) A sketch map of the nucleus boundary isoheights.
    The closer the isoheight deviates from the labeled nucleus boundary, the smaller the energy of nucleus boundary loss; The farther the isoheight deviates from the labeled nucleus boundary, the larger the energy of nucleus boundary loss.
    This boundary isoheight feature map can be used to approximate the closest distance between the estimated target boundary and the labeled boundary. Based on the approximated distance, we can penalize the deviations of the boundary estimations adaptively and efficiently.
    }
    \label{fig:Loss_nb}
\end{figure}

In order to comprehensively drive the learning of RepSNet, this paper integrates the optimization of the four feature maps NP, NT, BD, and NB together. The overall objective function is defined as follows:
\begin{equation}
\label{equ:L_total}
L = \lambda_{np} L_{np} + \lambda_{nt} L_{nt} + \lambda_{bd} L_{bd} + \lambda_{nb} L_{nb},
\end{equation}
where $L_{np}$, $L_{nt}$, $L_{bd}$ and $L_{nb}$ are the loss functions respectively for optimizing the feature maps NP, NT, BD and NB; And $\lambda_{np}$, $\lambda_{nt}$, $\lambda_{bd}$ and $\lambda_{nb}$ are the corresponding weight coefficients ( Four coefficients are set to 1 in this paper based on experiences and convenience considerations).

\vspace{0.25cm}
\noindent\textbf{(1) Nucleus foreground pixel loss function $L_{np}$ and nucleus type loss function $L_{nt}$:}

To learn the classification branch for NP feature map and NT feature map, we designed two loss functions based on cross-entropy (CE) and DICE coefficient \cite{GRAHAM2019HoverNet}. These two measures provide some complementary driving forces for the learning process. The DICE loss function enhances the learning ability of the model in the cases of imbalance between various classes, and the CE loss function ensures the stability of the learning. Therefore, the loss functions for learning NP feature map and NT feature map are defined as follows:
\begin{equation}
\label{equ:L_np}
L_{np} = \mathrm{CE}(\mathrm{NP}_g, \mathrm{NP}_p) + \mathrm{DICE}(\mathrm{NP}_g, \mathrm{NP}_p),
\end{equation}
\begin{equation}
\label{equ:L_nt}
L_{nt} = \mathrm{CE}(\mathrm{NT}_g, \mathrm{NT}_p) + \mathrm{DICE}(\mathrm{NT}_g, \mathrm{NT}_p),
\end{equation}
where $\mathrm{NP}_g(\mathrm{NT}_g)$ and $\mathrm{NP}_p(\mathrm{NT}_p)$
denote the labels and model estimates of the nucleus foreground (type), respectively;
$\mathrm{NP}_g =0$ indicates that the corresponding pixel does not belong to any nucleus, while {$\mathrm{NP}_g =1$} indicates that the corresponding pixel is a member of a nucleus.
In this work, the nucleus type label $\mathrm{NT}_g$ takes a value in the range from $0 \thicksim 6$ respectively for background, neutrophil, epithelial, lymphocyte, plasma, eosinophil and connective.

\vspace{0.25cm}
\noindent\textbf{(2) Nucleus boundary distance loss function} $L_{bd}$:

Due to the diversity of nucleus scale, the range of the distance from a nucleus pixel to the corresponding nucleus boundary differs from nucleus to nucleus.
Therefore, to ensure the prediction performance for BD feature map, we used the $\mathrm{Smooth}_{L1}$ loss \cite{girshick2015fast} for optimizing the regression branch. The corresponding loss function is defined as follows:
\begin{equation}
\label{equ:L_bd}
L_{bd} = \frac{1}{n}  \sum_{i=1}^n
   l(y_i, \hat y_i),
\end{equation}
where
\begin{equation}
\label{equ:L_bd_aux}
l(y_i, \hat y_i)=
    \begin{cases}
    0.5 (y_i - \hat y_i)^2, & \text{if } \mid y_i - \hat y_i \mid < 1 \\
    \mid y_i - \hat y_i \mid - 0.5, & \text{otherwise, }
    \end{cases}
\end{equation}
$\hat y$ and $y$ denote the reference label and model estimates of BD feature map, respectively.
Therefore, the $L_ {bd}$ measures the unreliability of the boundary distance feature map, and is referred to as a boundary distance loss function.

\vspace{0.25cm}
\noindent\textbf{(3) Nucleus boundary loss function} $L_{nb}$:

The PD method treats the discrimination of nucleus boundaries as a classification problem, and usually utilizes the CE loss to measure the inaccuracy of nucleus boundary estimation. In pathological images, however, typical phenomena are a widespread adhesion of nuclei and low contrast between cytoplasmic background and foreground of the nucleus. These phenomena result in the widespread existences of imperfect labels for nucleus boundary. Imperfect (anomalous) annotated boundaries make CE loss-based models difficult to be optimized and even impossible to converge. To reduce interferences from incorrect boundary labels, Zhou et al. \cite{zhou2019CIA-Net} proposed a smooth truncated loss. However, it is still limited to assessing the accuracy of boundary estimation probabilistically and fails to measure the distance discrepancy between the estimated boundary and the annotated boundary. This results in the same penalty regardless of how far the estimated nuclei boundary points deviate from the true boundary.

Therefore, to improve the accuracy and robustness of the nucleus boundary estimations from RepSNet, this work designed a deviation penalizing term specially for optimizing the NB feature map. For each estimated boundary point $p \in \mathrm{NB}$, suppose the shortest distance from $p$ to the labeled boundary is $d_p$. Then, the corresponding boundary loss is defined as $L_{nb} = \sum_{p\in \mathrm{NB}} d_p$. However, direct computation of $d_p$ is a task with a large space complexity and time complexity. To improve computational efficiency, we propose a method for approximately calculating the shortest distance based on a series of boundary isoheights. These boundary isoheights are generated by multiple morphological inflation from the manually annotated nucleus boundaries. Therefore, this distance is in fact the Chebyshev distance (the checkerboard distance).

Two columns of the subfigures in Fig. \ref{fig:Loss_nb} respectively represent the diagrammatic sketches of some nucleus instances and their corresponding boundary isoheights. For each nucleus instance, we plot its boundary isoheight feature map $\psi$ based on its manually labeled boundary. The closer an isoheight deviates from the labeled nucleus boundary, the smaller the corresponding energy of nucleus boundary loss. On the contrary
, the farther an isoheight deviates from the labeled nucleus boundary, the larger the corresponding energy of nucleus boundary loss. Therefore, we can approximate the shortest distance $d_p$ from an boundary pixel prediction $p$ to the labeled boundary by obtaining the position of this estimation on the isoheight feature map. The final boundary loss can therefore be rewritten as:
\begin{equation}
\label{equ:L_nb}
L_{nb} = \frac{\sum_{p\in\mathrm{NB}} \psi_p }{\tau \mid NB \mid + e},
\end{equation}
where $\tau$ denotes the maximum value of the isoheight ($\tau=5$ in this paper), $e$ is a smoothing constant, and $\mid NB \mid$ denotes the total number of pixels for estimating nucleus boundary. In equation (\ref{equ:L_nb}), $\sum_{p\in\mathrm{NB}} \psi_p $ is the cumulative errors between the estimated boundary and the reference boundary. This cumulative error is approximately calculated based on the boundary isoheights. The $\tau \mid NB \mid$ in the denominator makes $L_{nb}$ between 0 and 1, ensuring that this loss is comparable with the $L_{np}$, $L_{nt}$ and $L_{bd}$ in equation (\ref{equ:L_total}). The smoothing constant $e$ is to avoid a zero denominator in case of the NB degeneration. In conclusion, the $L_ {nb} $ measures the deviation of nucleus boundary estimation from its reference boundary, and is referred to as the nucleus boundary loss function.

\subsection{Connectivity Analysis and Nucleus Instance Classification}\label{subsec:connectivityAnalysisNucleusClassification}

We can detect the final nucleus instances by computing the connected areas using a connectivity analysis algorithm on the NB feature map inside the estimated nucleus boundaries. In addition, to prevent the loss of boundary information, we assign each predicted boundary pixel to its nearest nucleus instance. Finally, we count the number of occurrences of each category in every nucleus instance region, and select the category with the most occurrences as the final prediction result for the nucleus instance.

\section{Experiments}
\label{sec:Result}

This section systematically evaluate the performance of the proposed method in nucleus segmentation and classification. To this end, we first compare NBRIA(RepSNet) with some benchmark models on the Lizard dataset \cite{Graham2021Lizard}. This dataset is characterized by the large number of nuclei, many nucleus types, widespread nucleus adhesion areas. Therefore, the Lizard dataset is appropriate to be used for evaluating the segmentation performance and inference speed of the NBRIA(RepSNet). Then, we verified the effectiveness of the proposed components BVM, RepSNet, $L_ {nb} $ by some ablation experiments on NBRIA(RepSNet), more experiments conducted by individually using them in some benchmark models.

\subsection{Lizard Dataset}
\label{sec:Dataset}

All experiments in this work are conducted on the official Lizard \cite{Graham2021Lizard} dataset provided by the Colon Nuclei Identification and Counting Challenge (CoNIC) \cite{graham2021conic,CoNiC2023arXiv} in 2022. This dataset consists of a series of pathological images of colon tissues produced at 20x objective. These pathological images come from GlaS \cite{Korsuk2017glas}, CRAG \cite{GRAHAM2019MILD-Net}, CoNSeP \cite{GRAHAM2019HoverNet}, DigestPath \cite{DigestPath2022}, PanNuke \cite{PanNuke2019,gamper2020multi} and TCGA\cite{grossman2016toward}. To enable these pathological images derived from various data sources to be computed uniformly, Lizard proposed a multi-stage annotation pipeline. Using this pipeline, Lizard obtained approximately 500,000 labelled nuclei from H\&E-stained colon tissues. Such a large dataset of pathological images provides excellent supports for evaluating deep learning-based segmentation methods.

To ensure the consistency of the comparison experiments, this work utilizes the Lizard patches provided by the CoNIC organizer used for training and validation in this paper. The Lizard provides
 4981 $255*255*3$-sized H\&E image patches with corresponding mask labels and category labels (including neutrophil, epithelial, lymphocyte, plasma, eosinophil, and connective).  We randomly divided these patches into training, validation, and test sets in a ratio of 7:1:2, respectively, for model training, hyperparameter tuning, and comparative evaluation of model performance.

\subsection{Implementation Details and Evaluation Metrics}
\label{sec:Implementation}

This subsection describes the implementation details and evaluation metrics for NBRIA(RepSNet).

The NBRIA(RepSNet) model is trained and tested on a RTX 3090 GPU. To deal with the imbalance problem on the Lizard dataset, we conducted oversampling on the pathological images containing more nuclei from minority classes. During the training process, we employed geometric data augmentation and color augmentation techniques to improve the model's generalization capability. For optimizing model parameters, we chose the Adam optimizer with an initial learning rate 1×$10^{-4}$. In the cases without validation loss decrease for five consecutive training epochs, we halved the learning rate until it drops to less than 1×$10^{-7}$. It should be noted that NBRIA(RepSNet) does not depend on any pre-training based on other datasets, but only learns from the training set in section \ref{sec:Dataset}.

To comprehensively evaluate the performance of the NBRIA (RepSNet) model on the instance segmentation task, this paper employed multiple metrics. For example, the DICE score, Aggregated Jaccard Index (AJI)\cite{Kumar2017dataset}, and Panoptic Quality (PQ)\cite{Kirillov2019Panoptic}. The DICE score measures the degree of coincidence  between the model's predictions and the ground truth annotations. AJI focuses on evaluating the model segmentation quality for individual instances by computing the ratio of an aggregated intersection cardinality and an aggregated union cardinality. PQ comprehensively evaluates the model's performance on both semantic segmentation and instance segmentation. By comprehensively using these metrics, the model's performance can be evaluated objectively\cite{ZHAO2020Triple}.

The DICE score is defined as follows:
\begin{equation}
\label{equ:DICE}
\mathrm{DICE} = \frac{2 \times \mid X \cap Y \mid}{\mid X \mid + \mid Y \mid},
\end{equation}
where X and Y denote the masks for the predicted foreground and the ground truth foreground, respectively. $\mid ~ \mid$ represents the cardinality (number of elements) of a set.

AJI is a metric evaluating the nucleus instance segmentation performance by computing the intersection and union of predicted nuclei instances and ground truth nuclei instances \cite{GRAHAM2019HoverNet}.  The AJI is defined as follows:
\begin{equation}
\label{equ:AJI}
\mathrm {AJI} = \frac{\sum_{i=1}^n \mid G_i \cap P_{j_i} \mid} {\sum_{i=1}^n \mid G_i \cup P_{j_i} \mid + \sum_{k \in \varPsi} \mid P_k \mid},
\end{equation}
where $G=\{G_1, G_2, \cdots, G_n\}$ and $P = \{P_1, P_2, \cdots, P_m\}$ denote the sets of ground truth and predicted nucleus instance masks, respectively, $n$ is the total number of ground truth nuclei, and $m$ is the total number of predicted nuclei. ${j_i} = \arg\max\limits_k IoU(G_i, P_k)$, $IoU(G_i, P_k)={\frac{\mid G_i \cap P_k \mid } {\mid G_i \cup P_k \mid }}$, $ \varPsi = \{k: P_k \in P, \max\limits_{i=1,\cdots,n}IoU(G_i, P_k)  \leq 0 \}$.

Graham et al.\cite{GRAHAM2019HoverNet} advocated PQ as an indicator for nucleus instance segmentation to address the problem of excessive penalization in DICE and AJI on the cases in some overlapping regions. PQ is defined as follows:
{\footnotesize
\begin{equation}
\label{equ:PQ}
\mathrm{PQ} = \frac{2\mid TP\mid }{2\mid TP\mid  + \mid FP\mid  + \mid FN\mid } \times \frac{\sum\limits_{(x,y) \in TP} IoU(x, y)}{\mid TP\mid },
\end{equation}
}
where $x \in G$ and $y \in P$ denote the instance mask respectively for the labeled data and estimation results. $IoU(x,y) > 0.5$ indicates that $(x, y)$ is a pair of matched nuclei. Thus, $TP=\{(G_i, P_j): G_i \in G, P_j \in P, IoU(G_i, P_j) \ge 0.5 \}$ is the collection of the matched instance pairs. $FP=\{P_i: P_i \in P,$ and for any $G_j \in G, $ there is $(P_i,G_j) \notin TP \}$ denotes the set of estimated instances without any matched reference instances, $FN=\{G_i: G_i \in G,$ and for any $P_j \in P$,  there is $(G_i,P_j) \notin TP \}$ denotes the set of labeled instances without any matched prediction instances.
In addition, the multi-class PQ (mPQ) evaluation metric is utilized in this work to evaluate the performance of classification and segmentation synthetically.
The mPQ indicates that the PQ is computed independently for each positive type.

\subsection{Overall Evaluation of the NBRIA(RepSNet)}
\label{sec:comparison}


To evaluate the performance of the NBRIA(RepSNet) model, we compared it quantitatively (section \ref{sec:quantitative}) and qualitatively (section \ref{sec:qualitative}) with several open-source benchmark models, for example, Hover-Net \cite{ GRAHAM2019HoverNet}, DCAN \cite{chen2017DCAN}, U-Net \cite{Olaf2015UNet} and StarDist\cite{StarDist2018}. To deal with the problem of overlapping nuclei, DCAN\cite{chen2017DCAN} adds a contour detection branch on the top of U-Net, and it is shown that this scheme simultaneously obtains both instance segmentation and boundary determination for nuclei. To ensure fairness in comparison, both U-Net and DCAN use the same training dataset and training strategy as RepSNet in the experiments of this paper. In addition, Hover-Net \cite{GRAHAM2019HoverNet} is the official benchmark model provided by CoNIC and one of the best open-source models for nucleus instance segmentation. StarDist\cite{StarDist2018} is the State-of-the-Art method in CoNIC challenge. We retrained the Hover-Net and StarDist using the dataset in this work (Section \ref{sec:Dataset}).


\subsubsection{Quantitative analysis}
\label{sec:quantitative}

For a fair comparison, we used the code released by the authors for each benchmark model, conducted model training, validation and testing with the same dataset division (section \ref{sec:Dataset}). The experimental results are presented in Table \ref{tab:quantitative_analysis}. It is shown that the proposed NBRIA(RepSNet) scheme significantly outperforms other benchmark models on evaluation metrics such as AJI, Dice, PQ, and mPQ. These experimental results indicate the potential effectiveness and rationality of NBRIA(RepSNet) for the nucleus instance segmentation task. In particular, the instance segmentation performance is improved 1.61\% by the proposed NBRIA(RepSNet) compared with the SOTA model StarDist based on the mPQ. This improvement benefits from the innovative designs of the boundary voting mechanism and the boundary loss function, which enable the model to more effectively handle the segmentation of overlapping nuclei.

Additionally, Table \ref{tab:quantitative_analysis} presents the number of parameters (Params(M)), the number of floating-point operations per second (FLOPs), and the number of images that can be processed per second (Fps) for each model. Overall, although the NBRIA(RepSNet) improves the segmentation performance significantly, its efficiency is still very competitive, especially compared to the benchmark model Hover-Net. For U-Net and DCAN, although they have a smaller number of model parameters and a faster inference speed, their segmentation performance is significantly low due to the limitations of their model. For Hover-Net, in order to enable the model to learn horizontal-vertical distance maps, the authors used ResNet-101 as its backbone network, which resulted in Hover-Net being inferior to NBRIA(RepSNet) in terms of efficiency. StarDist achieved a relatively good balance between accuracy and inference speed. However, this model has certain limitations due to its reliance solely on the centroid pixels of each instance in predicting polygons. This limitation results in the lack of global contextual information in predicting nucleus boundary. In contrast, NBRIA(RepSNet), through structural re-parameterization techniques, maintains optimal segmentation performance while also achieving high inference efficiency. Consequently, NBRIA(RepSNet) significantly enhances the overall performance of pathological image instance segmentation, makes it feasible to achieve high-precision medical image segmentation on edge devices or embedded systems.

RepSNet not only exhibited excellent segmentation performance but also demonstrated some superiorities on training efficiency. Experiments show that  RepSNet converges faster than StarDist. Specifically, when the training strategy was the same and the number of training rounds was 30, the RepSNet's mPQ score on the test set is 0.538, while the StarDist is only 0.302. Additionally, when the training time was fixed at 5 hours, the mPQ score of RepSNet was 0.527, while that of StarDist was only 0.374. These results indicate that, regardless of whether the training epochs or training time were the same, RepSNet could achieve superior segmentation performance. The key to RepSNet's faster convergence speed lies in its innovative boundary prediction mechanism. Unlike StarDist, which predicts the entire nucleus boundary using individual pixels, RepSNet adopts an information fusion approach, utilizing the information from all pixels within a nucleus to jointly predict the boundary. This method can excellently guide model training, thus accelerate the convergence process. Consequently, with its outstanding segmentation performance and rapid training efficiency, RepSNet has exhibited remarkable potential in nucleus instance segmentation and classification tasks.

\begin{table*}[htbp]
    \centering
    \caption{Quantitative Evaluation of NBRIA(RepSNet) and Benchmark Models on the Test Set.}
    \label{tab:quantitative_analysis}
    \begin{tabular}{lccccccc}
        \toprule
        Method  & AJI    &   DICE    &   PQ      &   mPQ & Params(M) & FLOPs(B) &  Fps \\\hline
        U-Net\cite{Olaf2015UNet}  & 0.5182 &   0.8003  &   0.5054  &   0.4109 & 9.2 &   45.9 &   24\\
        DCAN\cite{chen2017DCAN} & 0.6360 & 0.8145 & 0.5901 & 0.4799 & 13.6 &  78.2 &  13  \\
        Hover-Net\cite{GRAHAM2019HoverNet} & 0.6629        & 0.8292     &  0.6275 &  0.5314  & 39.7 & 192.7   & 6 \\
        StarDist\cite{StarDist2018} & 0.6713        & 0.8365     &  0.6342 &  0.5472  & 14.7 & 92.4   & 12 \\
        NBRIA(RepSNet) & \textbf{0.6825} & \textbf{0.8413} & \textbf{0.6410}  & \textbf{0.5633} & 18.6 & 138.7 & 10 \\
        \bottomrule
    \end{tabular}
\end{table*}

\subsubsection{Qualitative analysis}
\label{sec:qualitative}

\begin{figure*}[htbp]
    \centering
    \includegraphics[width=1\linewidth]{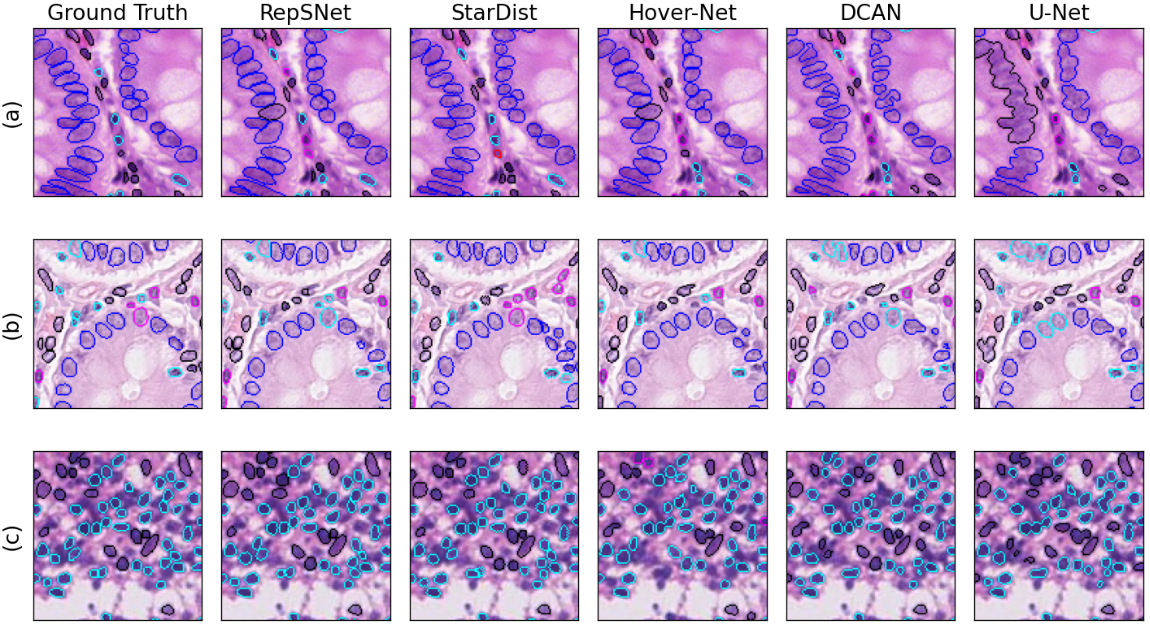}
    \caption{
     Qualitative comparisons between StarDist, Hover-Net, DCAN and U-Net methods. From left to right are the ground truth annotation (GT), segmentation results of NBRIA (RepSNet), StarDist, Hover-Net, DCAN, and U-Net. The cell nuclei boundaries in different colors represent different classes. NBRIA(RepSNet) demonstrates superior segmentation performance compared to other models, especially in areas with blurred boundaries (as shown in (a) and (b)) and areas with smaller cell nuclei (as shown in (c)). Even in these challenging cases, NBRIA (RepSNet) can segment individual cell nuclei instances very well. This further verifies the effectiveness and superiority of the proposed NBRIA (RepSNet) scheme for the task of cell nuclei instance segmentation.
    }
    \label{fig:Qualitative_analysis1}
\end{figure*}

\begin{figure*}[htbp]
    \centering
    \includegraphics[width=1\linewidth]{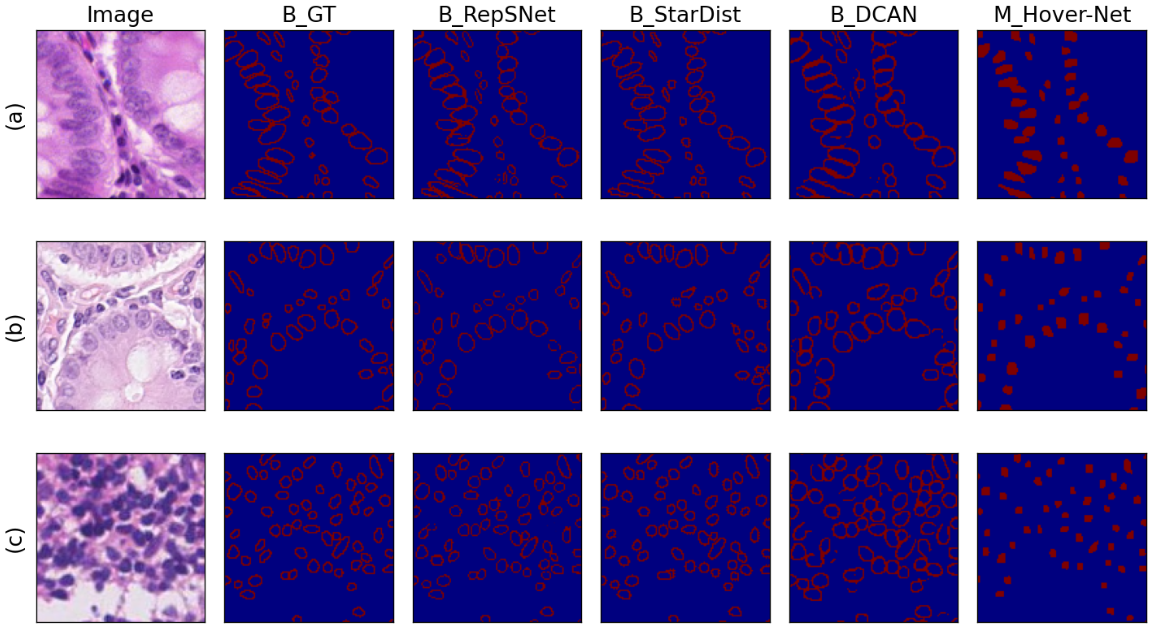}
    \caption{
    Feature maps of instance segmentation methods NBRIA(RepSNet), DCAN, and Hover-Net. This figure shows the input image, nucleus boundaries of ground truth (B\_GT), the nucleus boundary feature map B\_RepSNet estimated from NBRIA(RepSNet), the nucleus boundary feature map B\_StarDist estimated from StarDist, the nucleus boundary feature map B\_DCAN estimated from DCAN, and the marker feature map M\_Hover-Net estimated from Hover-Net from left to right. Compared to DCAN, NBRIA(RepSNet) excellently segment the nucleus instances even in areas with weakened or blurred nuclear boundaries. Unlike Hover-Net, NBRIA(RepSNet) can excellently recognize the targets and locate their boundaries even in areas with small nuclei.
    }
    \label{fig:Qualitative_analysis2}
\end{figure*}

To visually observe the instance segmentation performance of NBRIA(RepSNet), we visualized the instance segmentation results of this method and several representative benchmark models. Fig. \ref{fig:Qualitative_analysis1} presents some instance segmentation results of GT, NBRIA(RepSNet), StarDist, Hover-Net, DCAN, and U-Net methods respectively from left to right.
Overall, NBRIA(RepSNet) segments and recognizes the nucleus instances better than the other benchmark models. As U-Net only estimates the foreground regions of the nuclei, it cannot segment overlapping nuclei. This results in U-Net segmenting all of the connected pixels as one nucleus (Fig. \ref{fig:Qualitative_analysis1}(a)(b)).

\textbf{DCAN} adds a nucleus boundary estimation branch to U-Net and thus has some abilities to segment the adhesion nuclei.
However, the principle of DCAN is to directly estimate the outer boundary of a nucleus and segment the overlapping nuclei accordingly. Although this method has an excellent instance segmentation performance in areas with sparse nuclei, it can ``shrink'' the foreground targets in overlapping and dense areas (Fig. \ref{fig:Qualitative_analysis1} (b)). The ``shrink'' refers to the fact that the estimated nucleus instances are heavily missed at their boundaries. Different from DCAN, NBRIA(RepSNet) estimates the inner boundary for each nucleus and assigns the boundary pixel to the nearest nucleus instance after obtaining nucleus instances. Therefore, the NBRIA(RepSNet) does not ``shrink" the nucleus, even in areas with a high density of nuclei.

Hover-Net uses a horizontal-vertical distance estimation branch to separate overlapping nuclei. Although Hover-Net has excellent segmentation accuracy, it fails to deal with the cases with overlapping and small nuclei (see Fig. \ref{fig:Qualitative_analysis1}(c)).  While StarDist uses star polygons to approximately describe nucleus boundaries, StarDist only rely on the core region pixels of each instance to predict the nucleus boundaries. This characteristic leads to the lack of global contextual information for each polygon boundary prediction from StarDist. Therefore, the segmentation performance of StarDist was affected (Fig. \ref{fig:Qualitative_analysis1}(a) and (b)). In contrast, RepSNet uses a BVM-based post-processing scheme to determine the nucleus boundary. It is shown that the NBRIA(RepSNet) works well even in this kind cases with some overlapping and small nuclei.

In addition, this work further explores the differences on the instance segmentation performances between NBRIA(RepSNet) and PD methods (DCAN\cite{chen2017DCAN}), NCPID methods (Hover-Net\cite{GRAHAM2019HoverNet}) and StarDist\cite{StarDist2018} based on feature maps (Fig. \ref{fig:Qualitative_analysis2}). Fig. \ref{fig:Qualitative_analysis2} presents, from left to right, the input image (Image), the ground-truth nucleus boundary feature map (B\_GT), the estimated nucleus boundary feature maps from NBRIA(RepSNet) (B\_RepSNet), StarDist (B\_StarDist), DCAN (B\_DCAN), and Marker feature map from Hover-Net (M\_Hover-Net).

It is shown that DCAN is difficult to accurately identify nucleus boundaries in areas with overlapping nuclei, dense nuclei, or blurred nucleus boundaries. It is due to the fact that DCAN discriminates the identity of every pixel using local information without fusing any global features of any nucleus. This limitation results in the blurred boundary estimations in the above-mentioned cases for DCAN. NBRIA(RepSNet) finds the nucleus boundaries using a scheme based on a  macroscopic, integrated mechanism. This mechanism exploits and makes full use of the information in distances from each nucleus pixel to its parent nucleus boundary. As a result, NBRIA(RepSNet) provides more accurate nucleus boundary identification than PD-based methods such as DCAN. On the other hand, Hover-Net firstly obtains the CA markers of a nucleus from the horizontal-vertical distance map, and then uses a watershed algorithm based on these markers to segment each nucleus instance. Therefore, the identification accuracy of CA markers of nuclei directly determines whether the obtained nucleus instance segmentation results are complete. However, Hover-Net can not accurately identify the CA markers for some small nuclei. This inaccuracy leads to its omission of such small nuclei. Different from DCAN and Hover-Net, StarDist predict nucleus boundaries using star-convex polygons. However, its boundary prediction only depends on some local information around a single pixel and fails to effectively fuse the contextual information of the nucleus. Therefore, there are large deviations in the nucleus detection results of StarDist. In contrast, NBRIA(RepSNet) uses a boundary voting mechanism and multi-scale feature fusion scheme to exploit contextual information; Therefore, it can not only accurately detect small nuclei, but also make more accurate predictions for nucleus boundaries.


\subsection{\textbf{NBRIA(RepSNet) Interior Ablation Studies}}\label{sec:Ablation}

\begin{table*}[htbp]
    \centering
    \caption{Influences of key techniques on NBRIA(RepSNet).}
    \label{tab:Ablation}
    \begin{tabular}{cccccc}
        \toprule
        RepVgg &  RepUpsample  &  $L_{nb}$  &   AJI & DICE & mPQ \\\hline
        \ding{56} & \ding{56} & \ding{56} & 0.6517 & 0.8137 & 0.5324 \\
        \ding{56} & \ding{56} & \ding{52} & 0.6530 & 0.8172 & 0.5376 \\
        \ding{52} & \ding{56} & \ding{56} & 0.6748 & 0.8315 & 0.5537  \\
        \ding{52} & \ding{52} & \ding{56} & 0.6793 & 0.8364 & 0.5612 \\
        \ding{52} & \ding{52} & \ding{52} & \textbf{0.6825} & \textbf{0.8413} & \textbf{0.5633} \\
        \bottomrule
    \end{tabular}
\end{table*}

Interior ablation evaluation is to assess the effects of various key technologies by switching off individual technilogical option in NBRIA (RepSNet). In this subsection, we evaluate the effectiveness of the proposed scheme from three aspects: RepVgg Unit structure, RepUpsample(RepUp) structure, and nucleus boundary loss function $L_{nb}$. Table \ref{tab:Ablation} presents the performance evaluation results of several variants of the NBRIA(RepSNet) with various module  configurations. It is shown that NBRIA(RepSNet) is still able to achieve excellent instance segmentation performance even without any of the above-mentioned three techniques. This phenomenon indicates that it is indeed a feasible scheme to conduct nucleus instance segmentation through a NBRIA-based scheme. In case of replacing the traditional residual modules and upsampling modules with re-parameterizable RepVgg Units and RepUpsample modules, NBRIA(RepSNet) not only can synthetically exploit the effective features with various receptive fields, but also improves the inference efficiency. Additionally, the boundary loss function $L_{nb}$ can flexibly guide the prediction for nucleus boundaries. Therefore, the NBRIA(RepSNet) with the nucleus boundary loss function $L_{nb}$ improves the accuracy of nucleus boundary detection and overall instance segmentation performance.

\subsection{Exterior Ablation Studies based on Some Benchmark Nucleus Instance Segmentation Schemes}

Exterior ablation evaluation refers to the effectiveness assessing for the proposed key technology components by grafting them into some benchmark nucleus instance segmentation schemes. This subsection evaluated the effectiveness of the proposed boundary voting mechanism  (BVM), RepSNet network architecture, and the nucleus boundary loss function ($L_{nb}$) based on three benchmark nucleus instance segmentation schemes DCAN\cite{chen2017DCAN}, Hover-Net\cite{GRAHAM2019HoverNet} and StarDist\cite{StarDist2018}.

\subsubsection{Evaluation of BVM}

In this experiment, we first evaluated the impact of the boundary voting threshold on nucleus instance segmentation performance. It is showed that the optional configuration is 3 based on mPQ metric. Therefore, we consistently set the voting threshold to 3 in subsequent experiments.

To evaluate the effectiveness of BVM, we replaced the post-processing methods in Hover-Net and StarDist with BVM under the same encoder-decoder framework for comparison. Related experimental results are presented in Table \ref{tab:BVM}. It is shown that the BVM method significantly improved the nucleus instance segmentation performance on the whole based on metrics mPQ and DICE. Specifically, BVM improved the AJI score by 1.3\% and 0.66\%, the DICE score by 0.75\% and 0.32\%, the PQ score by 0.42\% and 0.39\%, and the mPQ score by 0.96\% and 0.71\%, respectively, based on Hover-Net and StarDist. These results further indicate the superiority of BVM over other distance regression-based post-processing methods in determining instance boundaries.

\subsubsection{Evaluation of RepSNet}

To evaluate the effectiveness of the proposed RepSNet network architecture for nucleus instance segmentation, we investigated it based on benchmark schemes Hover-Net and StarDist, which achieved the best performance on the Lizard dataset.

We indirectly evaluate the rationality of RepVGG and ReuUpsample by replacing the encoder and decoder in HoverNet and StarDist with the corresponding modules of RepSNet. Table\ref{tab:BVM} presents the experimental results on the Lizard dataset. It is shown that the proposed RepSNet module and RepVGG, RepUpsample improved Hover-Net and StarDist consistently. Specifically, compared to the corresponding original schemes (Hover-Net, StarDist), the variants improve them by (0.43\%, 0.69\%), (0.38\%, 0.28\%), (0.22\%, 0.40\%), and (0.41\%, 0.69\%) based on the AJI, DICE, PQ, and mPQ metrics, respectively. Moreover, it is worth mentioning that the inference speed for RepSNet is also significantly faster than that of Hover-Net and StarDist.  This further demonstrates the excellence of the proposed RepSNet model structure in the nucleus instance segmentation.

\subsubsection{Evaluation of the $L_{nb}$ Loss}

To evaluate the effectiveness of the proposed nucleus boundary loss $L_{nb}$, we introduced it into two benchmark models with boundary prediction capabilities, StarDist and DCAN. The experimental results are presented in Table \ref {tab:BVM}. It is shown that the $L_{nb}$ improved two benchmark models consistently. Specifically, compared with the original configuration, $L_{nb}$ improves StarDist by 0.18\% on AJI, 0.06\% on DICE, 0.2\% on PQ, and 0.4\% on mPQ. Compared with the original configuration, $L_{nb}$ improves DCAN by 0.42\% on AJI, by 0.5\% on DICE, 0.58\% on PQ, and  0.45\% on mPQ. These results indicate that the introduction of $L_{nb}$ contributes to  the nucleus segmentation and classification performance improvements. The improvement effects of $L_ {nb} $ are attributed to the accurate description of the deviation of the boundary prediction from reference. For the boundary predictions with large deviations, $L_ {nb} $ gives a large penalty to encourage the model to converge towards an accurate direction.

\begin{table*}[htbp]
    \centering
    \caption{Quantitative Evaluation of the Effectiveness of the Boundary Voting Mechanism (BVM), Network Architecture (RepSNet), and Boundary Loss ($L_{nb}$).}
    \label{tab:BVM}
    \begin{tabular}{ccccccccc}
        \toprule
        Model & $RepSNet_{encoder}$ & $RepSNet_{decoder}$ & Post-pro & $L{nb}$ & AJI    &   DICE    &   PQ      &   mPQ \\\hline

         \multirow{2}{*}{DCAN} & \ding{56} & \ding{56}  & Original   & \ding{56} &  0.6360  &  0.8145 &  0.5901 &  0.4799\\
         & \ding{56} & \ding{56}  & Original   & \ding{52} & \textbf{0.6402} &  \textbf{0.8195} &  \textbf{0.5959} & \textbf{0.4844}\\\hline

         \multirow{5}{*}{HoverNet} & \ding{56} & \ding{56}  & Original   & \ding{56} &  0.6629  &  0.8292 &  0.6275 &  0.5314\\
         & \ding{56} & \ding{56}  & BVM   & \ding{56} & \textbf{0.6759} &  \textbf{0.8367} &  \textbf{0.6317} & \textbf{0.5410}\\
         & \ding{52} & \ding{56}  & Original   & \ding{56} &  0.6653  &  0.8314 &  0.6283 &  0.5332\\
         & \ding{56} & \ding{52}  & Original   & \ding{56} &  0.6659  &  0.8327 &  0.6291 &  0.5353\\
         & \ding{52} & \ding{52}  & Original   & \ding{56} &  0.6672  &  0.8330 &  0.6297 &  0.5355\\\hline
        \multirow{6}{*}{StarDist}& \ding{56} & \ding{56}  & Original   & \ding{56} &  0.6713  &  0.8365 &  0.6342 &  0.5472\\
         & \ding{56} & \ding{56}  & BVM   & \ding{56} &  0.6779  & \textbf{0.8397} &  0.6381 &  \textbf{0.5543}\\
         & \ding{56} & \ding{56}  & Original   & \ding{52} &  0.6731  &  0.8371 &  0.6362 &  0.5512\\
         & \ding{52} & \ding{56}  & Original   & \ding{56} &  0.6764  &  0.8393 &  0.6373 &  0.5535\\
         & \ding{56} & \ding{52}  & Original   & \ding{56} &  0.6751  &  0.8384 &  0.6369 &  0.5524\\
         & \ding{52} & \ding{52}  & Original   & \ding{56} &  \textbf{0.6782}  &  0.8393 &  \textbf{0.6382}  &  0.5541\\
        \bottomrule
    \end{tabular}
\end{table*}

\subsection{Overall Evaluations based on Lizard Official Configurations}

Table \ref{tab:OnlineTest} presents a leaderboard of models for nucleus segmentation and classification tasks with mPQ scores exceeding 0.400\cite{CoNiC2023arXiv}. Models are ranked in descending order based on mPQ scores. It is shown that the epithelial cell, lymphocyte, and connective tissue cell classes are the easiest to segment, with mPQ scores of 0.556, 0.545, and 0.545, respectively. On the other hand, neutrophils and eosinophils are the most challenging to segment, with mPQ scores of only 0.281 and 0.330, respectively. This is due to the large class imbalance in the nucleus types on this dataset.

StarDist employs a geometric data augmentation method and a H\&E-based augmentation method during the data preprocessing phase. To deal with the imbalance problem on the nuclei from various types, StarDist oversamples the patches containing nuclei from the minority classes during training. At the final prediction phase, it integrates the results from multiple models with test-time augmentation (TTA). The third-ranked IFP Bern also utilizes geometric data augmentation, blurring data augmentation, noise augumentation,  H\&E-based augmentation, and oversamples the data from the classes with much less samples during training. It uses a weighted focal loss during training and multi-model ensemble for prediction. The fourth-ranked Pathology AI adds a large number of dropout layers in its up-sampling branch, uses a combination of Dice loss and weighted cross-entropy loss to enhance the model’s ability to deal with imbalanced data. Other models also adopt similar competitive strategies, for example, data augmentation, strategies handling class imbalance distribution and ensemble learning, etc.

In the basic edition of NBRIA(RepSNet), we only conducted some geometric data augumentation. The corresponding test result on CoNIC online system is 0.428 for mPQ (more results are presented in the 1st line of Table \ref{tab:ourlineTest}). The experimental results indicate that the key problem is underfitting. To this end, we conducted the following optimization: RepSNet network structure optimization and capacity improvement, data augumentation and oversampling, loss weight optimization, test-time augumentation, and dropout regularization, etc. For this optimized NBRIA(RepSNet), the CoNIC online test result is 0.478; off-line test result is 0.605. The off-line ensemble post-optimization investigations in the last paragraph of section \ref{sec:Result} indicate that NBRIA(RepSNet) has some possibilities and potentials to catch up with or exceed the State-of-the-Art StarDist with ensemble post-processing (unfortunately, the CoNIC test system was closed and related experiments cannot be conducted now).

\textbf{RepSNet Network structure optimization and capacity improvement:}
The encoder and decoder of RepSNet respectively consist of $n$ blocks, and each block consists of $m$ sub-modules (the sub-module in the encoder is a RepVGG, and the sub-module in the decoder is a RepUpsample). We tested various combinations for $n\in\{4, 5\}$ and $2\leq m\leq 12$, for example, [2,4,6,2] for $n=4$, [2,3,3,6,3] for $n=5$, [2,3,3,12,2]  for $n=5$ (the numbers in each square bracket are the choices for $m$). Based on experimental analysis, the configuration with $n=5$ and [2,3,3,6,3] is the optimal one.  The model with this configuration achieved the best mPQ score of 0.549 on the locally divided test set, which is an improvement of 2.3\% compared with the basic NBRIA(RepSNet) model. We retrained NBRIA(RepSNet) with this configuration based on the CoNIC official data set split. The CoNIC test results showed that the mPQ score increased from the original 0.428 to 0.440, and the $PQ^+$ scores for each category are presented in Table \ref{tab:ourlineTest}.

\textbf{Data Augmentation and Oversampling:}
The experimental results in Table \ref{tab:ourlineTest} shows that after optimizing the RepSNet network structure, the mPQ scores of many categories are improved, but the scores of the ``neu" and ``eos"  nuclei decrease. The main reason is that the proportion of "neu" and "eos" nuclei in the training data is very small, only accounting for 0.77\% and 0.63\%, respectively. The sample scarcity leads to severe insufficient feature learning for these two categories. To deal with this issue, we adopted data resampling and data augumentation strategies. For the "neu" and "eos" categories, we resampled the training images containing a relatively larger number of nuclei of these two categories to reduce the imbalance phenomenon. To prevent model overfitting, in addition to using conventional data augmentation methods such as geometric transformations, we also introduced new augmentation techniques like hue transformation and H\&E staining to increase the contrast between the foreground and background, and provided the model with more diverse training data. After the above-mentioned data resampling and data augmentation processings, the model's mPQ score on the local test set was improved to 0.552. Therefore, we retrained the enhanced model and submitted it to the official platform for test. The results showed that the mPQ score increased from 0.440 to 0.455, and the $PQ^+$ scores for each category are presented in Table \ref{tab:ourlineTest}. It is shown that the scores of the "neu" and "eos" categories have been significantly improved. The improvement indicates that the data resampling and data augmentation strategies effectively alleviate the class imbalance problem.

\begin{table*}[htbp]
    \centering
    \caption{Segmentation and classification challenge results on the final test set}
    \label{tab:OnlineTest}
    \begin{tabular}{ccccccccc}
        \toprule
       Model & $mPQ^+$ & $PQ^+_{Epi}$ &  $PQ^+_{Lym}$ &  $PQ^+_{Pla}$ &  $PQ^+_{Neu}$  &  $PQ^+_{Eos}$ &  $PQ^+_{Con}$\\\hline

        EPFL(Stardist) & 0.501 & 0.607 & 0.592 & 0.520 &  0.361  &  0.398 &  0.529 \\
        NBRIA(RepSNet) & 0.478 & 0.574 & 0.567 & 0.490 &  0.357  &  0.388 &  0.493 \\
        IFP Bern & 0.476 & 0.579 & 0.572 & 0.486 &  0.292  &  0.415 &  0.513 \\
        Pathology AI & 0.463 & 0.544 & 0.554 & 0.445 &  0.338  &  0.402 &  0.496 \\
        LSL000UD & 0.463 & 0.591 & 0.559 & 0.452 &  0.306  &  0.375 &  0.494 \\
        AI-medical & 0.458 & 0.566 & 0.568 & 0.485 &  0.282  &  0.343 &  0.501 \\
        Arontier & 0.457 & 0.571 & 0.554 & 0.491 &  0.259  &  0.361 &  0.507 \\
        CIA Group & 0.451 & 0.595 & 0.536 & 0.463 &  0.263 &  0.341 &  0.507 \\
        MAIIA & 437 & 0.572 & 0.541 & 0.421 &  0.252  &  0.351 &  0.484 \\
        ciscNet & 0.429 & 0.550 & 0.503 & 0.417 &  0.275  &  0.343 &  0.490\\
        MBZUAI\_CoNIC & 0.421 & 0.523 & 0.526 & 0.440 &  0.243  &  0.344 &  0.449\\
        Denominator & 0.408 & 0.508 & 0.544 & 0.427 &  0.213  &  0.294 &  0.460\\
        Softsensor\_Group & 0.405 & 0.466 & 0.539 & 0.429 &  0.275  &  0.283 &  0.434\\
        BMS\_LAB & 0.401 & 0.534 & 0.471 & 0.406 &  0.216  &  0.325 &  0.453\\
        \bottomrule
    \end{tabular}
\end{table*}

\begin{table*}[htbp]
    \centering
    \caption{Results of the NBRIA (RepSNet) on the CoNIC Challenge final test set}
    \label{tab:ourlineTest}
    \begin{tabular}{cccccccccc}
         \toprule
        Model & Strategy & $mPQ^+$ & $PQ^+_{Epi}$ &  $PQ^+_{Lym}$ &  $PQ^+_{Pla}$ &  $PQ^+_{Neu}$  &  $PQ^+_{Eos}$ &  $PQ^+_{Con}$\\\hline

         $A$ & Basic & 0.428 & 0.532 & 0.534 & 0.412 &  0.277  &  0.333 &  0.479 \\
         $A_1$ & Deepen Network & 0.440 & 0.562 & 0.551 & 0.470 &  0.249  &  0.317 &  0.489 \\
         $A_2$ & Oversampling+H\&E augment& 0.455 & 0.565 & 0.562 & 0.471 &  0.290  &  0.351 &  0.492 \\
         $A_3$ & Loss weight adjustment& 0.461 & 0.564 & 0.555 & 0.478 &   0.313  &  0.365 &  0.491 \\
         $A_4$ & TTA & 0.471 & 0.569 & 0.554 & 0.489 &   0.340  &  0.383 &  0.492 \\
         $A_5$ & Dropout & 0.478 & 0.574 & 0.567 & 0.490 &   0.357  &  0.388 &  0.493 \\
         \bottomrule



    \end{tabular}
\end{table*}

\textbf{Loss Weight Optimization:}
After data augmentation and oversampling, the model's segmentation performance for the minority classes "neu" and "eos" has been improved to some extent. However there is still an evident gap compared with the other classes. To further mitigate the issue of class imbalance, we adjusted the weights of the loss function to push the model's optimization process focusing on the features of the minority classes "neu" and "eos". Specifically, we modified the weight coefficients in the Focal Loss to increase the importance of these two classes in the classification loss. This adjustment makes the model more sensitive to the features of these two classes. Additionally, we adjusted the weight parameters of various tasks in the overall loss function to highlight the importance of boundary detection and nucleus classification. After adjusting the weights in the loss function, although there was no improvement in the model's mPQ score on the local test set (0.542), both the PQ+ scores for the "neu" and "eos" classes increased. These performance changes indicate that the adjustment strategy helps to improve the segmentation performance for these two minority classes. Therefore, we retrained the model with the adjusted loss weights and submitted it to the official platform for testing. The results showed that the mPQ score increased from 0.455 to 0.461, and the PQ+ scores for all categories are shown in Table \ref{tab:ourlineTest}.

\textbf{Test-Time Augmentation (TTA) Adjustment:}
According to the boundary voting mechanism proposed in this paper, the higher the proportion of correctly predicted boundary pixels, the better the model's segmentation performance. To improve the accuracy of boundary prediction, we enhanced the Test-Time Augmentation (TTA) strategy. Specifically, we extended the original method with four augmented transformations and predictions for each input image during the test phase to an updated edition with eight augmented transformations and predictions. `Augmented transformations and predictions' refers to first generate multiple sub-samples from each input image (such as flipping, rotation, etc.), then predict an instance segmentation result from each sub-sample, and fuse the predicted results from all sub-samples to obtain more accurate boundary prediction. It should be noted that this adjustment only affects the prediction mode of the test phase, without the need to retrain the model. We directly use the pre-trained model weights and perform prediction test according to the new TTA strategy. After the above-mentioned adjustments, the model's mPQ score on the local test set increased from 0.552 to 0.596, an increase of 4.4\%. These experiments indicate that the enhanced TTA strategy indeed improves the model's segmentation performance. Therefore, we applied the new TTA strategy to test the trained models and submitted the test results to the official platform. The results show that the mPQ score increased from the 0.461 to 0.471. The $PQ^+$ scores for each category are shown in Table \ref{tab:ourlineTest}. By enhancing the TTA strategy during test, the model's prediction for nucleus boundaries becomes more accurate. Therefore, this scheme improved the overall quality of segmentation and further enhanced the performance on the official platform.

\textbf{Dropout Regularization Technique:}
The above-mentioned experimental results show that, although a high mPQ score of 0.596 was achieved on the local test set, the mPQ score on the official test set is only 0.47126, indicating a significant gap between the two results. This discrepancy is probably due to overfitting, where the model performs much better on the training set than on new test data. To mitigate the overfitting issue, we further adopted a regularization technique, Dropout, to some feature maps. Specifically, at some levels of the model, we introduced Dropout, which randomly discards some information of the feature maps, thereby enhances the model's generalization ability. After the Dropout adjustment, although the model's mPQ score on the local test set (0.563) did not further increase, we retrained the model with the added Dropout and submitted its prediction results to the official platform for test. The results showed that the mPQ score improved from the original 0.47126 to 0.47808, the $PQ^+$ scores for each category are shown in Table \ref{tab:ourlineTest}.
From Table \ref{tab:ourlineTest}, it can be seen that the model does indeed exhibit a certain degree of overfitting. By using the feature map Dropout regularization strategy, the model's generalization ability is enhanced, and its segmentation performance on the official test set is improved.

\textbf{Ensemble Learning (tested off-line):}
To further improve the model's generalization ability and segmentation accuracy, we adopted a model ensemble strategy. After the official test channel was closed, we fused the prediction results from the four models A2, A3, A4, A5 in Table \ref{tab:ourlineTest}. This fusing procedure fully leverages the advantages of different models to compensate for the segmentation biases that may exist in a single model. Another benefit of model ensemble is that, by integrating multiple models that perform differently on the training data, we can reduce the risk of overfitting. A single model is prone to overfitting on the training data, but an ensemble of multiple models can mitigate this risk and improve the model's generalization ability. On the local test set, the model ensemble achieved a mPQ score of 0.605, with the PQ+ scores for each category as follows: "con" 0.641, "eos" 0.482, "epi" 0.725, "lym" 0.683, "neu" 0.466, "pla" 0.630. Compared with a single model, the ensemble model improved its mPQ score by 0.042. These experimental results indicate that model ensemble is an effective method to enhance segmentation performance on this task. Based on the excellent segmentation performance on the local test set, we have reason to believe that if the final ensemble model is submitted to the official test platform, there is some possibilities that test results is superior to the StarDist model, or comparable to it.

\section{Summaries and outlooks}
\label{sec:Summary}

This paper proposes a novel segmentation scheme for nucleus instances based on nucleus boundary regression and information aggregation (NBRIA), and designs a fully structural re-parameterizable network RepSNet based on this scheme.
RepSNet adopts a dual-branch encoder-decoder structure to predict the category of nuclei \& identity of each pixel, and the distance information of each pixel from its parent nucleus's boundary, respectively. The `identity' refers to whether a pixel comes from nucleus foreground or cytoplasmic background.

To compute accurate nucleus boundary prediction based on the extracted boundary distance information, we propose a synergistic belief enhancement scheme as a post-processing pipeline. This synergistic belief enhancement scheme is implemented by a designed method referred to as the boundary voting mechanism (BVM). This mechanism fuses the distance information of a series of pixels to their parent nucleus's boundaries, enhances the robustness of the proposed NBRIA(RepSNet).

Moreover, to improve the computational efficiency of the model and its sensitivity to multi-scale features,
numerous re-parameterizable modules are designed in RepSNet. During the training phase, these modules enable the method to learn the multi-scale feature sensibility by some receptive fields with various scales. During the inference phase, a re-parameterization technique is adopted to reduce the number of parameters and computational burdens by merging multi-scale feature extraction branches. To evaluate the performance of the proposed NBRIA(RepSNet), we conducted various experiments on the Lizard dataset \cite {Graham2021Lizard}. The experimental results shown that NBRIA(RepSNet) achieves superior segmentation accuracy and faster inference speed than some benchmark models such as StarDist\cite{StarDist2018} and Hover-Net\cite{GRAHAM2019HoverNet}. Additionally, we submitted our proposed scheme to the CONIC challenge platform\footnote{CONIC challenge platform:\url{https://conic-challenge.grand-challenge.org/}} for test, and achieved the second place in the cell nuclei instance segmentation and classification task. StarDist, the State-of-the-Art method on CoNIC challenge, used an ensemble learning post-processing procedure, and the NBRIA(RepSNet) didn't utilized this strategy. However, the CoNIC test system of the platform has been closed, and further online test cannot be submitted. Fortunately, some off-line ensemble post-optimization investigations was conducted and its experimental results indicate that NBRIA(RepSNet) has some possibilities and potentials to catch up with or exceed the State-of-the-Art StarDist with ensemble post-processing. Additionally, a series of experiments based on off-line data indicates the superiorities of the proposed NBRIA(RepSNet) over the benchmark methods, for example, StarDist, Hover-Net, and DCAN.

The proposed scheme is designed based on the idea of nucleus boundary regression and information aggregation. This scheme fully exploits the local information and global\&contextual information. A series of experiments were conducted to evaluate its effectiveness and superiorities over benchmark methods in the scenarios of nucleus instance segmentation for pathological images. In future, the scheme can be further optimized and investigated for more applications in other medical image segmentations or natural image segmentations. With the increasing of medical images, NBRIA(RepSNet) can also be improved by designing more directions for estimating the  distances from each pixel to its parent nucleus's boundary. In theory, more distance directions help to improve the representation capability of the model, and potentially increase the segmentation accuracy. However, this adjustment inevitably increases the model complexity, and needs more data for learning. In general, the scheme is reasonably designed and has the potential for expansion, and can be explored and optimized in more scenarios in the future.


\bmhead{Acknowledgments} This work is supported by the National Natural Science Foundation of China (Grant No. 12373108, 11973022), the Natural Science Foundation of Guangdong Province (No. 2020A1515010710).

\bmhead{Data and code availability}  The experimental code, dataset splitting configuration and the pre-trained model  were released at  \url{https://github.com/luckyrz0/RepSNet}. 




\bmhead{Conflict of interest} We have no known conflicts of interest.

\begin{appendices}



\end{appendices}


\bibliography{reference}


\end{document}